\documentclass[aps,reprint,twocolumn,showpacs,superscriptaddress]{revtex4-1}
\usepackage{amsmath, amssymb, braket, dsfont, times}
\usepackage{graphicx}
\usepackage{epstopdf}
\usepackage[breaklinks]{hyperref}

\usepackage[usenames,dvipsnames]{color}			
\definecolor{Zcolour}{rgb}{0.992, 0.588, 0.22}
\definecolor{dkgreen}{rgb}{0,0.5,0}
\definecolor{purple}{rgb}{0.5,0,0.5}

\newcommand{\bea}{\begin{eqnarray}}
\newcommand{\eea}{\end{eqnarray}}
\newcommand{\bpm}{\begin{pmatrix}}
\newcommand{\epm}{\end{pmatrix}}

\newcommand{\imth}{\hspace{1pt}\mathrm{i}\hspace{1pt}}

\newcommand{\bse}{\boldsymbol{e}}
\newcommand{\se}{U}  
\newcommand{\sg}{\textrm{SG}} 
\newcommand{\pse}[1]{G_{#1}} 
\newcommand{\psg}{\textrm{PSG}} 
\newcommand{\igg}{\textrm{IGG}} 
\newcommand{\pso}[1]{\hat{G}_{#1}\hat{#1}}
\newcommand{\psoinv}[1]{({\hat{G}_{#1}\hat{#1}})^{-1}}
\newcommand{\bc}{\textrm{bc}}

\begin{document}

\title{Measuring  space-group symmetry fractionalization in Z$_2$ spin liquids}
\author{Michael P. Zaletel}
\affiliation{Department of Physics, Stanford University}
\author{Yuan-Ming Lu}
\affiliation{Department of Physics, University of California, Berkeley, California 94720, USA}
\author{Ashvin Vishwanath}
\affiliation{Department of Physics, University of California, Berkeley, California 94720, USA}

\begin{abstract}
The interplay of symmetry and topological order leads to a variety of distinct phases of matter, the Symmetry Enriched Topological (SET) phases. Here we discuss physical observables that distinguish different SETs in the context of Z$_2$ quantum spin liquids with SU(2) spin rotation invariance. We focus on the cylinder geometry, and show that ground state quantum numbers for different topological sectors are robust invariants which can be used to identify the SET phase. More generally these invariants are related to 1D symmetry protected topological phases when viewing the cylinder geometry as a 1D spin chain. In particular we show that the Kagome spin liquid SET can be determined by measurements on  one ground state, by wrapping the Kagome in a few different ways on the cylinder. In addition to guiding numerical studies, this approach provides a transparent way to connect bosonic and fermionic mean field theories of spin liquids. When fusing quasiparticles, it correctly predicts nontrivial phase factors for combining their space group quantum numbers. 

\end{abstract}
\maketitle
\tableofcontents
\section{Introduction}
In contrast to conventional phases that are distinguished by Landau order parameters, topologically ordered states with emergent anyonic excitations remain distinct even in the absence of symmetry. However, the presence of symmetries, which is natural in most physical contexts, leads to further distinctions, the so called symmetry enriched topological phases (SETs). Well known manifestations include fractional charge of anyons in the fractional Quantum Hall states (and in fractional Chern insulators) and spin-charge separation in quantum spin liquids. Recently, rapid progress in the theoretical understanding of SETs is being made \cite{Wen2002, Kitaev2006,MesarosRan, EssinHermele2013, LuAV2013,HungWan2013,AVSenthil2013,FidkowskiLindnerKitaev,BurnellChenFidkowskiAV,Barkeshli2014,SongHermele}. This is partly driven by conceptual advances in the related theory of strongly interacting Symmetry Protected Topological phases (SPTs),  where despite the absence of anyon excitations in the bulk, nontrivial edge excitations emerge \cite{ChenGuWen2011,PollmannTurnerBergOshikawa2010, Schuch,FidkowskiKitaev2011,ChenSPT,LevinGu,LuAV2012,AVSenthil2013}. What would be particularly welcome at this stage, is a physically well motivated example of an SET. To make progress in this direction we will need to understand  how different SET phases can be distinguished.

Further motivation for studying measurable characteristics of  SETs is the recent progress in the search for quantum spin liquids in frustrated magnets, both in experiments and in numerics. A number of S=1/2 quantum magnets in 2D and 3D frustrated lattice have been identified, which appear to evade magnetic order \cite{PALee,Balents}. These include the S=1/2 Kagome material, Herbertsmithite, which shown no sign of ordering down to temperatures that are a thousand times smaller than the exchange constant \cite{Helton2007}. While clear cut evidence of an energy gap in these materials is yet to emerge, a requirement to be considered topologically ordered, this may be an extrinsic effect due to impurities  (although other explanations have also been suggested \cite{Ran2007,Sachdev2010,Cepas2008}). Meanwhile, extensive density matrix renormalization group (DMRG) simulations of the nearest neighbor Kagome Heisenberg antiferromagnet (KHA) indicate (i) a gapped ground state that respect all symmetries \cite{Yan2011} and (ii) a topological entanglement\cite{Jiang2012,Depenbrock2012} entropy of $\log 2$. In reference \cite{MikeAV2014} it was argued that the ground state must possess Z$_2$ (toric code) topological order to be  compatible with (i) and (ii). An important open question is the identification of the precise phase of matter, i.e. the SET, realized by the Kagome antiferromagnet. While a complete solution would necessitate extensive numerical input, and is beyond the scope of this paper, here we will relate SET phases to physical properties that are readily measurable in numerical simulations.  

A prerequisite is a measurement of the topological order itself. Entanglement entropy \cite{LevinWenTEE,KitaevPreskill2006} provides one signature, although it does not uniquely specify the topological order. A complete characterization is obtained, from either the entanglement spectrum  \cite{LiHaldane} ( in certain cases), or the S and T matrices  \cite{Wen1993, Zhang2012, Wen, ZhangGroverAV, ZaletelMongPollmann2013, TuZhangQi, Haah}  some of which are well suited to numerical calculations \cite{Vidal,ShengHaldane, Gong2014,Bauer2014}. Related techniques can be used to diagnose 2D SPT phases \cite{Mike2014}.  

Here we discuss measurable properties that distinguish SETs. For the reasons above we focus on the S=1/2 KHA, assuming Z$_2$ (toric code) topological order, which has a pair of  emergent S=1/2 spinons (one bosonic and one fermionic) and a vortex (vison).
The different SETs differ in their realization of space group symmetries and their interplay with time reversal.

We consider  systems on both finite and infinite cylinder geometries, which are well suited to DMRG calculations.
We show that 1) the many-body quantum numbers of a finite-cylinder ground state under space group symmetries such as reflection, translation etc.  provide a powerful diagnostic of the underlying SET  and 2) when viewing  an infinitely long cylinder as a 1D spin chain, the 1D SPT order of the spin chain for various geometries and topological sectors completely determines the SET order, at least within the space of mean-field parton ansatz.

A reflection quantum number is a probe of quantum entanglement: if a state is odd under reflection, then the two halves of the system are entangled.
But how can quantum numbers of a global symmetry, which are always integral, be related to symmetry fractionalization?
As a simple example - consider creating a pair of identical anyons from the vacuum. Say that the pair is related to one another by a symmetry (such as a reflection or rotation). If the excitations carry charge, a unit charge for this state actually implies half charge for each excitation since they are constrained by symmetry.  This is one of the key ideas that we will exploit. Its implementation is more involved when the second symmetry is not charge, but also a space group (or time reversal) symmetry. Nevertheless such arguments establish the relative quantum numbers between different topological sectors. 

To relate our results to an established classification scheme, we use a specific model of SETs obtained by a parton decomposition of the spin operator into bosonic (Schwinger bosons) or fermionic (Abrikosov fermions) partons.
Symmetry fractionalization is encoded in the Projective Symmetry Group (PSG) \cite{Wen2002,WangAV, Huh2010}, which determines how the partons transform under symmetry. A parton mean field theory combined with projection leads to a spin wave-function whose quantum numbers in each topological sector are completely determined, which reflect the underlying SET.   In particular we show that the Kagome spin liquid SET can in principle be uniquely determined by measurements on  {\em one} ground state in a few different finite-cylinder geometries.
This information is numerically superior to the relative quantum numbers between topological sectors, since  DMRG numerics on KHA do not obtain all topological sectors in the finite sized systems studied.

A different perspective on our approach is to view it as a `dimensional reduction' in which we view a 2D SET phase on a cylinder  as a 1D SPT.
The nature of the 1D SPT depends on the topological sector being studied, the geometry of the cylinder, and the SET. 
In addition to its utility as a diagnostic in numerics, our approach provides a theoretical tool to study connections between different representations of the same SET. 
The 1D SPT invariants for the degenerate ground states (which are labeled by quasiparticles of the topological order) are shown to follow the same multiplicative law as the fusion rules for the Abelian quasiparticles.  
This allows us to determine from the PSGs of two anyons types in the Z$_2$ topological order the PSG of the third anyon type, which is found to obtain nontrivial phases in certain cases.
Some of these were not previously known \cite{EssinHermele2013} and serve to correctly relate bosonic and fermionic mean field states on the Kagome lattice \cite{LuAV2013}.
In particular this leads us to equate two popular states, the $Q1=Q2$ Schwinger boson state \cite{Sachdev1992} and the Z$_2[0,\pi]\beta$ fermionic mean field state \cite{LuRanLee,LuAV2013,HongYao}. 

In addition, we show how dimensional reduction can be used to completely identify the four topological sectors of a cylinder;
this is  highly useful for DMRG studies, and does not require simultaneous knowledge of all four ground states.
In particular, we have found a 1D SPT invariant that distinguishes between the bosonic and fermionic spinon.

Earlier work employed a similar dimensional reduction approach  in the case of internal symmetries with projective representations, \cite{PollmannChen} and here we find it to be much more generally applicable in the presence of space group symmetries. Related work  specializing to the case of just translation symmetry has recently appeared \cite{Wang2014}.  The connection between 1D SPT invariants and global quantum numbers was previously noted \cite{PollmannTurnerBergOshikawa2010, TurnerZhangAV}, and other works have utilized global many body quantum numbers to identify topological phases \cite{Rigol,Ran2014}. 

\section{Review of Z$_2$ spin-liquids}
	According to the arguments of Hastings, Oshikawa, Lieb, Shultz and Mattis, \cite{LSM1961,Oshikawa2000,Hastings2005} a quantum magnet with half-integer spin per unit cell is either gapless, breaks a symmetry, or is a gapped `spin - liquid' with emergent anyonic excitations. 
In the latter case, the simplest possibility consistent with time-reversal is the Z$_2$ `toric code'-type spin-liquid.  \cite{MikeAV2014} 
This phase has emergent $S = 1/2$ excitations, the `spinons,' even though a truly local excitation (the magnons) must carry  $S = 1$.
These emergent spinons are anyonic excitations with non-trivial braiding and statistics.
The Z$_2$ spin-liquid has four anyon types: the local excitations (`1'); the bosonic spinon (`$b$'), which carries $S = 1/2$; the vison (`$v$'), which behaves like a $\pi$-flux for the spinon; and the fermionic spinon (`$f$'), formed from the composite of $f = b v$, and which also carries $S = 1/2$.
Each particle is its own anti-particle, $v^2 = f^2 = b^2 = 1$ (hence  `Z$_2$').
The braiding and statistics of the Z$_2$ spin-liquid are equivalent to Z$_2$ gauge theory (the `toric code').
In the language of Z$_2$ gauge theory, $b$ is the electric charge $e$; $v$ is the magnetic flux $m$; and $f$ is the dyon $f = e m$ composed of flux and charge.

The simplest SET aspect of the Z$_2$ spin-liquid is its behavior under SO(3) spin rotations:
$1$ and $v$ carry integer spin, while the spinons $b, f$ carry half-integer spin.
The half-integer spin carried by the spinons is `fractionalized' because any local excitation of the constituent $S = 1/2$ spins transforms as $S = 1$.
We can look for additional SET distinctions based on the transformation properties of anyons under space-group symmetries, which is the subject of this work.

\subsection{Minimally entangled states}
	The Z$_2$ spin liquid has a  4-fold topological ground state degeneracy on the torus or cylinder. 
Throughout this work we rely on a special basis for the ground-state manifold  called the `minimally entangled states' (MES).\cite{Zhang2012}
To construct the MES basis, let $x$ run along the infinite length of the cylinder, and let  $\mathcal{F}_x^a$ denote the adiabatic process in which  a pair of anyons $a / \bar{a}$ are created from the vacuum and dragged in opposite directions $\pm \hat{x}$ out to infinity.
The process $\mathcal{F}_x^a$ returns the system to the ground state, so is a unitary operation in the ground-state manifold.
We say that $\mathcal{F}_x^a$ `threads anyonic flux $a$' through the cylinder.
Since the model is Abelian, $\mathcal{F}_x^a \mathcal{F}_x^b \propto \mathcal{F}_x^{a \cdot b}$, where $a \cdot b$ denotes the  fusion of Abelian anyons.

The MES basis is the unique basis in which $\mathcal{F}_x^a$ is realized as a \emph{permutation} of the basis states  for all $a  \in \{  1, b, v, f \}$.
$\mathcal{F}_x^a$ acts as a permutation in the MES basis  because each MES has definite topological flux threading the cylinder (its `topological sector').

	In previous discussions of the MES it is  often assumed that each of the four MES can be uniquely identified with an anyon type $1, b, v, f$, so that each MES can be labeled with an anyon type $\ket{a}$.
Then threading topological flux is realized as $\mathcal{F}_x^b \ket{a} = \ket{b \cdot a}$.
This is the case for Z$_2$ spin-liquids on \emph{even} circumference cylinders, but for an $S = 1/2$ model on an odd circumference cylinder there is a subtlety that arises because the MES double the unit-cell along the $x$-direction.
The unit cell doubles because  $S = 1/2$ models behave as if there is a  topological flux piercing each unit cell, so (as for a magnetic field) the net topological flux through the cylinder changes along the direction $x$. \cite{MikeAV2014}
Strictly speaking it is more precise to view the MES as a torsor for the fusion group, but this subtlety does not affect  the measurements proposed in this work, so for notational simplicity we will label the MES by anyon types.

\section{Space-group quantum numbers: robust SET invariants}
	In this section we consider the space-group quantum numbers of  a finite length cylinder: even though the structure of the two edges is non-universal, we argue that the space-group quantum numbers are.
In particular, consider creating a  pair of well-separated excitations from the vacuum which are related by a mirror plane, and separating them out to the edges of the cylinder.
If the reflection quantum number flips sign after this process,  the excitations must be anyons which are connected by an invisible string which is \emph{odd} under reflection.
Since the pair of anyons together transforms as $R = -1$, it as if $R = \sqrt{-1}$ acting on each anyon individually, so we say $R$ is `fractionalized.'

While  global quantum numbers are only well defined for finite systems, at a later point we will show that they leave their imprint on the bulk entanglement spectrum in a way which can be measured on an infinitely long cylinder as well.
	
\subsection{Reflection quantum numbers}
\label{sec:reflectionQN}
	Consider a large but finite cylinder with a reflection symmetry $\hat{R}_x$ that exchanges its two edges.
We argue that the global  $\hat{R}_x$ quantum number of any symmetric state $\ket{\Lambda, a}$ with no excitations in the interior of the cylinder depends only on a) the SET order of the bulk phase of matter b) the details of the geometry, such as its dimensions, which we denote by `$\Lambda$',  and c) the topological sector $a$ of the cylinder.
Throughout this paper we will denote these global quantum numbers by $Q$,
\begin{align}
\hat{R}_x \ket{\Lambda, a} = Q_{R_x}(\Lambda, a)  \ket{\Lambda, a}.
\end{align}
To elaborate on c), note that an \emph{infinitely} long cylinder has the same topological ground state degeneracy as the torus.
While the edges may reduce the ground state degeneracy,  we only require that there are no excitations in the \emph{bulk} of the cylinder, so are left with the same bulk  degeneracy as the torus.
We label these topological sectors of the finite cylinder by $a$, and as discussed earlier we assume $a$ indexes a special `minimally entangled' basis which has definite topological flux $a$ threading the cylinder.
This work  focuses mainly on Z$_2$ topological order because the states $\ket{\Lambda, a}$ will \emph{break} $R_x$ if $a$ is not it's own anti-particle.

The global quantum number $Q_{R_x}$ is insensitive to any details of the edge state or bulk Hamiltonian.
To show insensitivity to the edge, note that perturbing the edges amounts to acting with unitaries $U_L, U_R$ localized at the edges of the cylinder.
When $U_L$ and $U_R$ are spatially well separated, using $R_x$ symmetry we can require that $U_R = \hat{R}_x U_L \hat{R}_x^{-1}$.
Since the perturbation $U_L \hat{R}_x U_L \hat{R}_x^{-1}$ commutes with $\hat{R}_x$, the quantum number is unchanged.
$Q$ is insensitive to the bulk Hamiltonian because $Q = \pm 1$, so is quantized and can only change during a bulk phase transition.

However,  $Q_{R_x}(\Lambda, a)$  \emph{can} depend on the topological sector $a$, since changing the topological sector from $a \to b \cdot a$  requires separating an anyon pair $b / \bar{b}$ out to the edges using $\mathcal{F}_x^b$,  which is a string-like operation.
We will find that the dependence on $\Lambda$ cancels if look at the relative quantum number between topological sectors, \footnote{Note that only the torsor structure of the MES is required here, which is why the procedure still applies to odd-circumference cylinders even though our notation is imprecise.}
\begin{align}
Q_{R_x}^{(b)} \equiv \frac{Q_{R_x}(\Lambda, b a) }{ Q_{R_x}(\Lambda, a) }.
\end{align}
These ratios have a particularly simple relationship to the SET order: if $Q^{(b)}_{R_x} = -1$, it implies that a pair of anyons related by $R_x$ each carry half the $R_x = -1$ quantum number, which we consider to be `fractional.'
In contrast, our earlier argument implies that a pair of truly local excitations must always have $Q_{R_x} = 1$.

	Now suppose there is an additional reflection symmetry $R_{y}$ which does \emph{not} exchange the two edges. In the absence of the edge-exchanging $R_x$ symmetry, the $R_y$ quantum numbers are not robust,  because nothing then prevents the perturbation $U_L$ from being odd under $R_y$ while $U_R$ is even.
But if both $R_x$ and $R_y$ are present, we can instead measure the combination $I = R_x R_y$; since $I$ exchanges the edges, according to our earlier reasoning $Q_I$ is also a protected invariant.

	Finally, if lattice doesn't have a $C_4$ symmetry (as for the Kagome model)  there are  distinct ways to compactify the geometry into a cylinder: for one choice, $R_x$  exchanges the edges, while  for the second, $R_y$ does.
We can then measure quantum numbers $Q_{R_x}$ in the first cylinder, $Q_{R_y}$ under the second cylinder, and $Q_{I}$ in either.	
This give three independent quantum numbers for each anyon type, $Q^{(b / f / v)}_{R_x / R_y / I}$, which we will find almost fully characterizes Z$_2$ SETs (at least within the PSG framework).
The remaining information relates to the commutation relations of time-reversal $\mathcal{T}$ and the reflections $R$, which will lead to protected edge degeneracies we discuss in Sec.\ref{sec:1D_RT}.

\subsection{Translation quantum numbers}
\label{sec:translationQN}
	In a magnetic field the translations $T_x, T_y$ form a `magnetic algebra'  $T_x T_y T_x^{-1} T_y^{-1} = e^{i \Phi} $ which is a \emph{projective} representation of the translation group.
Even in the absence of a physical magnetic field, in a topologically ordered phase the anyons may experience an effective magnetic field.
The magnetic field experienced by anyon $a$ is encoded in the projective relation $(T_x T_y T_x^{-1} T_y^{-1} )^{(a)} = \eta_{xy}^{(a)}$.

	For an Abelian theory the projective relations must obey the fusion rule $\eta_{xy}^{(a)} \eta_{xy}^{(b)} = \eta_{xy}^{(a b)}$, since $\eta^{(a)}_{xy}$ is  the Berry phase acquired when $a$ circles a unit cell.
For a Z$_2$ spin-liquid in an $S = 1/2$ model, we always have the relation $\eta_{xy}^{(v)} = -1$. 
This can be argued in the language of Z$_2$ gauge theory, where all $S = 1/2$ objects are the source of Z$_2$ electric flux (for example the spinons $b, f$, which map on to the electrically charged $e, f$ particles in the gauge theory).
This includes the microscopic $S = 1/2$ in each unit cell, which implies that the system behaves as if there is electric flux piercing each unit cell.
Consequently under $e$-$m$ duality, the flux $m$ (the vison) experiences a background flux of $\pi$ per unit cell, so  $\eta_{xy}^{(v)} = -1$.
Since $\eta_{xy}^{(1)} = 1, \eta_{xy}^{(v)} = -1$, and $\eta_{xy}^{(b)} \eta_{xy}^{(v)} = \eta_{xy}^{(f)}$, there is a single sign left undetermined, which is the most basic SET distinction between Z$_2$ spin-liquids.

	To probe $\eta_{xy}$, consider a cylinder of length $L_x$ and circumference $L_y$ in topological sector $a$,  and measure the momentum quantum number $Q_{T_y}(\Lambda, a) = \hat{T}_y \ket{\Lambda, a}$.
Note that when $L_y$ is odd the MES double the unit cell in the $x$ direction, so we must restrict to $L_y$ even.
$T_y$ symmetry alone does not protect $Q_{T_y}$, since the edge excitations can carry an arbitrary momentum, but the combination of $T_x$ and $T_y$ allows us to define a robust `momentum per unit length.'
Recall that in the Landau gauge, the momentum $T_y$ of a particle in a magnetic field is proportional to its position $x$.
Since $\ket{\Lambda, a}$ has an $a$ particle localized near the edge, as we grow $L_x$ its momentum  grows linearly with $x$, ie, it is a momentum per unit length of cylinder.
To define the momentum per unit length operationally, we need to grow the length of the cylinder $L_x$ while keeping the  topological flux and edge state the same, as otherwise $Q_{T_y}$ could contain a spurious contribution coming from the changing edge state.
Concretely, we require the reduced density matrix for the edge be kept constant as $L_x$ grows.
The momentum per unit length $\eta^{(a)}_{xy}$ is then
\begin{align}
Q_{T_y}(\Lambda, a) = q(\partial \Lambda, a)  \left( \eta^{(a)}_{xy} \right)^{L_x}.
\end{align}
$q(\partial \Lambda, a)$ depends on the edge, while the bulk contribution reveals the SET invariant $\eta^{(a)}_{xy}$.

The momentum per unit length is \emph{trivial} to measure in any tensor network ansatz.
For MPS, it is the invariant $\eta^{(a)}_{xy} = r_{T_y}$ defined in Eq.~\eqref{eq:mps_onsite}) when viewing $g = T_y$ as an `onsite' symmetry in the 1D representation of the cylinder.
$r_{T_y}$ is a  byproduct of the algorithm used to calculate the momentum-resolved entanglement spectrum of infinite-DMRG studies, so presumably has already been computed in existing studies.

\section{Symmetry enriched order: detecting the Projective Symmetry Group}
The preceding discussion is independent of any classification of space-group SETs, since we have argued on general grounds that these quantum numbers are robust SET invariants.
Nevertheless, we would like to identify these invariants within a general classification scheme. 
Currently there is not a complete classification of space-group SETs.
However, the parton construction provides a rich zoo of Z$_2$ spin-liquids whose symmetry properties can be analyzed using Wen's `projective symmetry group' (PSG). \cite{Wen2002}
The PSG provides at least a partial classification of space-group SETs. 
In this section, we show how to compute the quantum numbers $Q_{\se}(\Lambda, a)$ within the parton construction,  thereby identifying  them with invariants of the PSG.

\subsection{The Parton Construction and the Projective Symmetry Group}	

The resonating valence bond (RVB) picture proposed by Phil Anderson provided the  first intuition for a spin liquid ground state as a quantum superposition of different dimer configurations  covering a lattice of spin-$1/2$ particles.
Each dimer (denoted by $\bullet-\bullet$) is a singlet pair formed by two spin-$1/2$ particles:
\bea
|\bullet-\bullet\rangle=|\uparrow\rangle_1|\downarrow\rangle_2-|\downarrow\rangle_1|\uparrow\rangle_2.\notag
\eea
The state is a `liquid' because the quantum superposition of dimer patterns restores the translational symmetry.
One type of elementary excitation in these systems is created by breaking  a dimer into a pair of particles carrying spin-$1/2$ each, which were coined \emph{spinons}.

The \emph{parton} construction is a systematic formalism for writing down ansatz RVB wavefunctions  in which each spinon is realized either as a fermionic parton $f_\sigma$ or bosonic parton $b_\sigma$, where $\sigma=\uparrow,\downarrow$.
In the fermionic description, each dimer is realized as an s-wave Cooper pair of partons; breaking a Cooper pair generates a pair of spinons.
The microscopic spins $\vec S_{\bf r}$ are related to the partons through the bilinears
\bea\label{eq:parton:spin operators}
\vec S_{\bf r}=\frac12\sum_{\alpha,\beta=\uparrow,\downarrow}f^\dagger_{{\bf r},\alpha}\vec\sigma_{\alpha,\beta}f_{{\bf r},\beta}=\frac12\sum_{\alpha,\beta=\uparrow,\downarrow}b^\dagger_{{\bf r},\alpha}\vec\sigma_{\alpha,\beta}b_{{\bf r},\beta}.
\eea
where $\vec{\sigma}$ are the three Pauli matrices.
In order for this mapping to generate a sensible $S = 1/2$ wavefunction, the partons cannot be free particles: they obey the ``single-occupancy'' constraint of one parton per lattice site:
\bea
\sum_{\sigma=\uparrow,\downarrow}f_{{\bf r},\sigma}^\dagger f_{{\bf r},\sigma}=\sum_{\sigma=\uparrow,\downarrow}b_{{\bf r},\sigma}^\dagger b_{{\bf r},\sigma}=1,~~\forall~\text{lattice site}~{\bf r}.
\eea
This constraint can be implemented by a gauge field which couples to the partons.

In practice, we use the parton construction to create ansatz wave-functions. 
If $\ket{\text{MF}}$ is a `mean-field ansatz' state for the partons which need not obey the single occupancy constraint (for example, a BCS superconductor of fermionic partons $f_\sigma$), we  enforce the constraint via Gutzwiller projection to obtain an $S = 1/2$ wavefunction:
\bea
&\notag\langle\uparrow_1\downarrow_2\cdots\sigma_i\cdots|\Psi\rangle=\\
&\langle0|f_{{\bf r}_1,\uparrow}f_{{\bf r}_2,\downarrow}\cdots f_{{\bf r}_i,\sigma_i}\cdots \ket{\text{MF}},
\label{eq:Gproj}
\eea
and similarly in the bosonic construction. Note that in the fermionic case,  we must choose and fix an ordering of sites ${\bf r}_1, {\bf r}_2, \cdots$ in order to maintain the correct relative sign between different spin configurations.
For the purposes of calculation $\ket{\text{MF}}$ is usually taken to be a free wavefunction, such as a mean-field BCS superconductor or pair-superfluid for the fermionic / bosonic constructions considered here.
Gutzwiller projecting the creation of a \emph{single} parton $f^\dagger / b^\dagger$ results in a highly non-trivial excitation: an $S = 1/2$ anyonic excitation, the spinon.

A crucial question in the parton  construction is how the symmetries $\{ \se \in \sg\}$ of the $S = 1/2$ wavefunction (such as global SO(3) spin rotations and space-group symmetries) are realized in the partons and their mean-field ansatz $\ket{\text{MF}}$.
The simplest possibility is that the partons form a linear representation of the symmetry group $\sg$.
But is this the only option?
The answer is no, because according to Eq.~\eqref{eq:parton:spin operators}  applying a gauge transformation $b_{{\bf r},\sigma} \to e^{\imth \phi_{\bf r}} b_{{\bf r},\sigma}$ leaves the physical spin operators  unchanged, and accordingly the Gutzwiller projection in Eq.~\eqref{eq:Gproj} is a many-to-one mapping insensitive to $U(1)$ gauge transformations.
Consequently under a series of symmetry operations $\{ \se_i\} \in \sg$ which yield the identity operation $\bse$
\bea
\se_1 \se_2\cdots \se_n=\bse\label{eq:parton:series of symmetry}
\eea
a single-parton operator $b_{{\bf r},\sigma}$ (or $f$) may acquire a nontrivial phase factor $e^{\imth \phi} \neq1$ instead of remaining invariant.
In this case the partons transform \emph{projectively}, rather than linearly, under the symmetry group $\sg$.
The symmetry operations $\{ \se \in \sg\}$ are accompanied by certain gauge transformations $\{ \pse{\se}|\se\in\sg \}$ on partons, forming a ``projective symmetry group'' $\psg \equiv \{\pse{\se}\se|\se \in \sg\}$ which is  a central extension of the symmetry group $\sg$. \cite{Wen2002} 
The center of such an extension is called the ``invariant gauge group'' $\igg  \equiv \psg / \sg $.
The $\igg$ are those gauge transformations which leave the mean-field ansatz $\ket{\text{MF}}$ invariant.

For the Z$_2$ spin liquids which are the focus of this work, the invariant gauge group is $\igg = Z_2$. This means under a series of symmetry operations $\{ \se_i\}$ in (\ref{eq:parton:series of symmetry}), each parton acquires a $Z_2$ phase factor of $\pm1$. This is consistent with the mean-field ansatz being a singlet BCS superconductor of spinons.
For example, for symmetry a symmetry $U U = \bse$ we have the relation
\bea
(\pso{U})^2 f_{{\bf r}_i,\sigma_i}(\pso{U})^{-2} = \eta^f_{U}f_{{\bf r}_i,\sigma_i},~~~\eta^f_{U}=\pm1.
\eea
The primary goal of this work is to show how to detect these $\pm 1$ phases associated with symmetry group $\sg$.

\subsection{Minimally entangled states within the parton construction}
	We first must review how to generate the MES within the parton construction.
	
	There is a two-fold degeneracy associated with threading a vison $v$ for both  finite and infinite cylinders.
In the parton ansatz this arises because   the boundary conditions of a cylinder of circumference $L_y$ can be either periodic (P) or anti-periodic (AP).
At the level of the parton Hamiltonian, this is accomplished by assigning an additional sign of $-1$ to all matrix elements which cross a line at some fixed $y = y_0$; the choice of $y_0$ is a gauge choice. 
When acting with $T_y$ or $R_y$, the location of the twist $y_0$ must be restored by an additional contribution to the gauge transformations $\pse{T_y / R_y}$, leading to new PSG relations:
\begin{align}
\label{eq:bc_psg}
( \pso{T_y} )^{L_y} &=  (-1)^{\bc} \\
( \pso{T_y} )^{L_y / 2}(\pso{R_y}) (\pso{T_y} )^{-L_y/2}  \psoinv{R_y} &= (-1)^{\bc}
\end{align}
where $\textrm{bc} = 0 / 1$ for $P / AP$.
Hence in all that follows the PSG implicitly depends on the boundary condition ($\bc$) of the mean-field ansatz $\ket{\textrm{MF}}$	

	The two-fold degeneracy associated with threading a spinon is a bit more subtle, as in the finite case the degeneracy is split by the edges.
If we make a bipartition of the cylinder at some $x_0$, the parton parity $(-1)^{N_{b / f}}$ in the left half of the mean-field ansatz $\ket{\textrm{MF}}$ fluctuates across the cut (note the parton number itself is not conserved).
After Gutzwiller projection, Eq.~\eqref{eq:Gproj}, the parton parity to the left is fixed by the number of sites to the left.
But in the infinite case there is an ambiguity, since the number of sites is infinite.
This means that when Gutzwiller projecting we can freely choose either the sector with even (E) or odd (O) parton parity to the left of the cut at $x_0$, which generates an additional 2-fold degeneracy on the infinite cylinder.

	The choice of P / AP boundary conditions combined with E / O parton parity generates the 4-fold degeneracy of the infinite cylinder. These sectors are identified with the anyon types in a manner that depends on the parton construction:
\begin{align}
1, v, b, f &\leftrightarrow  \mbox{(P, E), (AP, E),  (P, O), (AP, O)} \quad (\mbox{bosonic}) \\
1, v, b, f &\leftrightarrow  \mbox{(AP, E), (P, E),  (AP, O), (P, O)} \quad (\mbox{fermionic})
\end{align}	
Note the role of P/AP is flipped between the two constructions.

For an even circumference cylinder the E / O parity is the same for all cuts $x_0$, since an even number of sites intervene between cuts.
But for an odd circumference cylinder, the E / O assignment alternates with $x_0$.
This alternation doubles the physical unit cell.

\subsection{Computation of global quantum numbers from the PSG}
Since the PSG determines how the partons transform under symmetry operations, we can compute the (crystal) symmetry quantum numbers of any projected wavefunction (Eq.~\eqref{eq:Gproj}) constructed from a parton mean-field ansatz.
If $\se$ is a space-group operation which permutes the sites according to $i \to \se(i)$, the wavefunction transforms as
\begin{align}
\label{eq:gutz_transform}
\braket{  \{ \sigma_i \}  | \se \Psi } & =  \braket{  \{ \sigma_{\se^{-1}(i)} \}  | \Psi }=  \bra{0} \prod_i f_{{\bf r}_i, \sigma_{\se^{-1}(i)}} \ket{ \text{MF}  },
\end{align}
and similarly for the bosonic ansatz. We can split the resulting quantum number into two contributions. First, there is a part $Q_U(\Lambda, \bc)$ coming from the Gutzwiller projection,
\begin{align}
\label{eq:proj_transform}
\bra{0} \prod_i f_{{\bf r}_i, \sigma_{\se^{-1}(i)}} = Q_{U}(\Lambda, \bc)  \bra{0} \prod_i f_{{\bf r}_i, \sigma_{i}}  \pso{\se},
\end{align}
which depends only on the geometry $\Lambda$ and the PSG (which is modified by $\bc$); we will show how to compute this in the subsequent section.
The second contribution comes from the quantum number of $\ket{ \text{MF}  }$.
Inserting Eq.~\eqref{eq:proj_transform} into Eq.~\eqref{eq:gutz_transform} we find that the total quantum number factorizes as
\begin{align}
\pso{\se}  \ket{ \text{MF}  } &\equiv   Q_U(\text{MF} )  \ket{ \text{MF}  } \\
\hat{U} \ket{\Psi } &= Q_{U}(\Lambda, \bc)  Q_U(\text{MF} ) \ket{\Psi } 
\end{align}

\subsection{Ratios of edge-exchanging quantum numbers in different topological sectors}
In the first scenario, we suppose we have access to all several topological sectors on the same geometry.
We find that for edge-exchanging symmetries, the ratio between the quantum number before and after threading anyonic flux $a$ reveals the PSG of anyon $a$.
\subsubsection{Spinon insertion}
	We fix the geometry $\Lambda$ and compute the relative reflection  quantum number $Q_R$ between two states  which differ by the insertion of a pair of spinons at the edges.
To generate the appropriate pair of mean-field ansatz $\ket{ \text{MF}  }$ and $\ket{ c \cdot \text{MF} }$ which differ by spinon flux $c = b / f$ (depending on the construction), let $c^\dagger_L$ create an arbitrary bosonic / fermionic parton near the left edge.
To ensure that $c \cdot \text{MF}$ is symmetric under $R$ we must create a corresponding spinon on the right using $c_R^\dagger \equiv ( \pso{R})  c^\dagger_L \psoinv{R}$, so
\begin{align}
\ket{ c \cdot \text{MF} } \equiv  c^\dagger_L  \, \,     (\pso{R}) c^\dagger_L \psoinv{R}  \ket{\text{MF}}.
\end{align}
$Q_{U}(\Lambda, \bc)$ is unchanged, and it is straightforward to verify
\begin{align}
& \pso{R} \ket{ b/f \cdot \text{MF}} &= (-1)^F ( \pso{R})^2  Q_R(\text{MF} ) \ket{ b/f \cdot \text{MF}}\end{align}
where the sign $(-1)^F$ occurs for fermionic partons as we must exchange the creation operators and we use $(\hat{\pse{R}} \hat{R})^2$ to denote the parton PSG associated with $\hat{R}^2=\bse$:
\bea
&\notag( \pso{R})^2 c_i ( \pso{R} )^{-2}=\eta^c_R~c_i,~~~\forall~c_i,\\
&(\pso{R})^2\equiv\eta^c_R.
\eea
Consequently for any geometry the change in the quantum number $Q_R$ after inserting a bosonic / fermionic spinon is
\begin{align}
Q^{(b/f)}_R \equiv \frac{Q_R(\Lambda,  b/f \cdot \text{MF})}{Q_R(\Lambda,  \text{MF})} &= (-1)^F \eta^{b/f}_R.
\end{align}
Hence the spinon PSG can be recovered by measuring the relative quantum number between different topological sectors.

The relative quantum numbers can be computed for any space-group symmetry which exchanges the edges; by using different cylinder compactifications we can measure the $R_x, R_y$, and $I = R_x R_y$ quantum numbers.
There may be distinct $\pi$-rotations depending on whether the rotation is site or bond / plaquette centered; on a square lattice, for example, $I' = T_x R_x T_y R_y$, reveals an independent PSG relation.
In Figs. \ref{fig:SSL_Qa} and \ref{fig:KSL_Qa} we tabulate the relative quantum numbers for the square and Kagome lattices.

\subsubsection{Vison insertion}
	We again fix $\Lambda$, and compute the relative reflection quantum number $Q_R$ between two states  which differ by the insertion of a vison. 
As discussed, threading a vison switches between P/AP boundary conditions, so $Q_{U}(\Lambda, \bc)$ may change due to the PSG relations of Eq.~\eqref{eq:bc_psg}.
The PSG relations associated with $U^2 = \bse$ are only modified if $U$ takes an odd number of sites across the twist boundary condition modified by the vison.
For both the square and Kagome lattice, for the geometries in which $U$ exchanges the edges of the cylinder we have
\begin{align}
Q_{R_y}(\Lambda, \textrm{AP}) &= Q_{R_y}(\Lambda, \textrm{P}) \\
Q_{R_x}(\Lambda, \textrm{AP}) &= Q_{R_x}(\Lambda, \textrm{P}) \\
Q_{I_{b/p}}(\Lambda, \textrm{AP}) &= Q_{I_{b/p}}(\Lambda, \textrm{P}) \\
Q_{I_{s}}(\Lambda, \textrm{AP}) &= -Q_{I_{s}}(\Lambda, \textrm{P}) 
\end{align}
where $I_{b/p}$ is a is bond or plaquette centered $\pi$-rotation, while $I_{s}$ it site-centered.
	
	Next we argue that $Q_U(\textrm{MF})$ is unchanged when threading a vison.
To start, suppose $\ket{\textrm{MF}}$ is in the same phase as the ground state of a BCS superconductor / pair super-fluid. 
In the subsequent section, we show that $Q_U(\textrm{MF}) = 1$ regardless of boundary condition, so is unchanged by vison insertion.
Now suppose we modify the ground state $\ket{\textrm{MF}}$ with arbitrary $U$-symmetric edge perturbations of fixed parton parity.
The resulting $Q_U(\textrm{MF})$ can depend only on the parton parity of the edge perturbation, but not on the boundary condition, because the vison-modified PSGs of Eq.~\eqref{eq:bc_psg} will necessarily act on $U$-related partons and the signs will cancel.
This shows $Q_U(\textrm{MF})$ is unchanged by vison insertion, and we conclude that
\begin{align}
Q^{(v)}_{R_y}&= 1, \quad Q^{(v)}_{R_x}&= 1, \quad Q^{(v)}_{I_{b/p}} &= 1, \quad Q^{(v)}_{I_{s}}&= -1.
\end{align}
A possible loophole in our argument is that we assumed that for the ground state, $\ket{\textrm{MF}}$ could be taken to be in the same phase as a BCS / pair superfluid, but our result agrees with earlier discussions.\cite{LuAV2013}

\subsubsection{Fusion and unification: $Q^{(f)} = Q^{(v)} Q^{(b)}$}
	Our derivation of the vison quantum numbers shows that the relative reflection quantum numbers obey fusion, $Q_U^{(f)} = Q_U^{(v)} Q_U^{(b)}$, because  
vison insertion changes the quantum number by $Q^{(v)}$ regardless of the parton parity at the edge. 
As tabulated in Table \ref{fig:SSL_Qa} and Table \ref{fig:KSL_Qa}, we can use this fusion rule to equate the bosonic and fermionic PSGs, unifying the two approaches.
\begin{table}
\begin{tabular}[t]{l || r r r | l}
$a$ & $b$ & $f$ & $v$ & $Q^{(f)} = Q^{(v)} Q^{(b)}$\\
\hline
$Q^{(a)}_{R_x}$ & $\eta_{R_x}^b$ & $-\eta_{R_x}^f $& 1 & $\eta_{R_x}^b = - \eta_{R_x}^f $\\
$Q^{(a)}_{R_y}$ &  $\eta_{R_y}^b$ & $-\eta_{R_y}^f$ & 1 &  $\eta_{R_y}^b = - \eta_{R_y}^f $ \\ 
$Q^{(a)}_{I_{p/b}}$ &  $\eta_{R_x}^b \eta_{R_y}^b  \eta^b_{R_x, R_y} $ & $-\eta_{R_x}^f  \eta_{R_y}^f  \eta^f_{R_x, R_y}$ & 1 & $ \eta_{R_x, R_y}^b = - \eta_{R_x, R_y}^f $\\
$Q^{(a)}_{I_s} $ & $\eta_{xy}^b \eta_{R_x}^b \eta_{R_y}^b  \eta^b_{R_x, R_y} $ & $-\eta_{xy}^f \eta_{R_x}^f  \eta_{R_y}^f  \eta^f_{R_x, R_y}$ & -1 & $\eta_{xy}^b = -\eta_{xy}^f $\\
\end{tabular}
\caption{Relative quantum numbers $Q^{(a)}_{\se}$ between topological sectors $a$ of a \textbf{square} lattice spin-liquid. $I_{p/b}$ is a plaquette or bond centered $\pi$-rotation, while $I_s$ is a site-centered $\pi$-rotation.}
\label{fig:SSL_Qa}
\end{table}
\begin{table}
\begin{tabular}[t]{l || r r r | l}
$a$ & $b$ & $f$ & $v$ & $Q^{(f)} = Q^{(v)} Q^{(b)}$\\
\hline
$Q^{(a)}_{R_x}$ & $({-1})^{p_2 + p_3}$ & $-\eta_{\sigma}$ & 1 & $(-1)^{p_2 + p_3} = - \eta_{\sigma} $\\
$Q^{(a)}_{R_y}$ &  $({-1})^{p_2}$ & $-\eta_{\sigma} \eta_{\sigma C_6}$ & 1 &  $({-1})^{p_3} = \eta_{\sigma C_6} $ \\ 
$Q^{(a)}_{I_h}$ &  $({-1})^{p_1 + p_3}$ & $-\eta_{C_6}$ & 1 & $ ({-1})^{p_1} = - \eta_{C_6} \eta_{\sigma C_6} $\\
$Q^{(a)}_{I_s} $ &  $({-1})^{p_3}$ & $-\eta_{12} \eta_{C_6}$ & -1 & $(-1)^{p_1} = -\eta_{12} $\\
\end{tabular}
\caption{Relative quantum numbers $Q^{(a)}_{\se}$ between topological sectors $a$ in a \textbf{Kagome} lattice spin-liquid. $I_{h/s}$ is a hexagon / site centered $\pi$-rotation.
The bosonic PSGs are expressed through the invariants $(p_1, p_2, p_3)$ of Ref.~\cite{WangAV}, while the fermionic PSGs are expressed through the invariants $\eta$ of Ref.~\cite{LuRanLee}.
The fermionic invariants  satisfy $\eta_{12}  \eta_{C_6} \eta_{\sigma C_6} \equiv 1$ tautologically.}
\label{fig:KSL_Qa}
\end{table}

\subsection{The quantum numbers of the ground state}
\label{sec:GS_QN}
A given geometry $\Lambda$ generically has a lowest energy state which is SO(3) symmetric on both edges, and two-fold degeneracy for vison insertion.
We can compute the quantum numbers for such an SO(3) symmetric state by assuming  $\ket{\text{MF}}$ is in the same universality class as a BCS superconductor / pair superfluid in the fermionic / bosonic constructions.
$\ket{ \text{MF}}$ is then invariant under any symmetry: $Q_U( \text{MF}) = 1$.
This is because in both constructions we have
\bea
&|\text{MF}\rangle=\exp\Big[\sum_{{\bf i},{\bf j}}g({{\bf i}-{\bf j}})c^\dagger_{{\bf i},\uparrow}c^\dagger_{{\bf j},\downarrow}\Big]|0\rangle,\\
&\notag g({\bf i}-{\bf j})= (-1)^F g({\bf j}-{\bf i})
\eea
where $c = f / b$ and $(-1)^F = -1 / 1$  for Abrikosov-fermions and Schwinger-bosons respectively.
Because $|\text{MF}\rangle$ always contains the parton Fock vacuum $\ket{0}$ as a component in the Taylor expansion of the exponential, and $\ket{0}$  is neutral under any symmetry operation, $\ket{\text{MF}}$ must also be neutral.
Therefore the quantum number of the ground state depends only on the geometry $\Lambda$ and the PSG:
\begin{align}
Q_U(\Lambda, \bc, \text{MF}) = Q_U(\Lambda, \bc)
\end{align}
where $\bc$ will depend on the two-fold degeneracy associated vison insertion and $Q_U(\Lambda, \bc)$ is defined in Eq.~\ref{eq:proj_transform}.
This result is particularly useful numerically, since by modifying $\Lambda$ we can probe the PSG using only a \emph{single} topological sector.

In the following we specifically illustrate how to obtain the eigenvalues $Q_U(\Lambda, \bc)$ of two crystal symmetry operators, inversion $I$ and mirror reflection $R$, for a projected wavefunction (\ref{eq:Gproj}) on a finite-size lattice from parton PSGs.

\subsubsection{Eigenvalue of plaquette-centered inversion $I_p$}

First we consider an inversion symmetry $I_p$ whose inversion center lies on a plaquette. For a finite-size lattice with plaquette-centered inversion $I_p$, the number of lattice sites $N_s$ must be even. All lattice sites must be exchanged in pairs under inversion operation, since no lattice site remains invariant under $I_p$ operation. More specifically, spinon $f_{{\bf r}_i,\sigma_i}$ must appear altogether with its inversion counterpart $f_{\hat{I}_p{\bf r}_i,\sigma_i}=\pso{I_p}f_{{\bf r}_i,\sigma_i}\psoinv{I_p}$ in the many-spinon operator $\prod_{i}f_{{\bf r}_i,\sigma_i}$. Note that in projected wavefunction (\ref{eq:Gproj}) there is always a particular ordering for the real-space positions $\{{\bf r}_i\}$ of the many-spinon operator $\prod_{i}f_{{\bf r}_i,\sigma_i}$. Here we simply choose a ordering in which a pair of spinons related by inversion show up together i.e.
\bea
&\prod_{i}f_{{\bf r}_i,\sigma_i}\equiv\prod_{i}^\prime f_{{\bf r}_i,\sigma_i}\cdot \pso{I_p}f_{{\bf r}_i,\sigma_i}\psoinv{I_p}.
\eea
where $^\prime$ denotes the product over half lattice sites that are unrelated by inversion. Clearly under inversion operation $\hat{I}_p$ the above many-spinon operator transform as
\bea
&\notag\pso{I_p}\prod_{i}f_{{\bf r}_i,\sigma_i}\psoinv{I_p}=\\
&\prod_{i}^\prime \pso{I_p}f_{{\bf r}_i,\sigma_i}\psoinv{I_p}\cdot (\pso{I_p})^2f_{{\bf r}_i,\sigma_i}(\pso{I_p})^{-2}.\notag
\eea
By definition of PSGs we have
\bea
(\pso{I_p})^2f_{{\bf r}_i,\sigma_i}(\pso{I_p})^{-2}=\eta^f_{I_p}f_{{\bf r}_i,\sigma_i},~~~\eta^f_{I_p}=\pm1.
\eea
since $I_p^2=\bse$ yields the identity operation. We used $\eta^f$ to denote the PSGs for fermionic spinons (Abrikosov fermions) and $\eta^b$ for bosonic spinons (Schwinger bosons). Notice that for Abrikosov-fermion representation, exchange of two spinons $f_{{\bf r}_i}$ and $f_{\hat{I}_p{\bf r}_i}$ gives rise to an extra $-1$ sign due to Fermi statistics. As a result we obtain
\bea
\pso{I_p}\prod_{i}f_{{\bf r}_i,\sigma_i}(\pso{I_p})^{-1}=(-\eta^f_{I_p})^{N_s/2}\prod_{i}f_{{\bf r}_i,\sigma_i}
\eea
Hence the eigenvalue of plaquette-centered inversion $I_p$ for projected wavefunction (\ref{eq:Gproj}) on a $N_s$-site lattice ($N_s=$even) is
\bea\label{eq:inversion:plaquette:f}
Q_{I_p}(N_s)=(-\eta^f_{I_p})^{N_s/2}
\eea
for Abrikosov-fermion representation. On the other hand, in the Schwinger-boson representation, exchange of two spinons won't yield a $-1$ sign and we have
\bea\label{eq:inversion:plaquette:b}
Q_{I_p}(N_s)=(\eta^b_{I_p})^{N_s/2}
\eea

\subsubsection{Eigenvalue of site-centered inversion $I_s$}

Now let's take one more step to consider an inversion symmetry $I_s$ whose inversion center lies on one or more lattice sites. Let's assume inversion centers contain $N_I$ sites and $N_s$ is the total number of lattice sites. For those $(N_s-N_I)$ sites other than the inversion centers, their contribution to the $I_s$ eigenvalue follows exactly the same form as (\ref{eq:inversion:plaquette:f}) and (\ref{eq:inversion:plaquette:b}), except that we need to replace $N_s$ by $N_s-N_I$.

What about the contribution from the $N_I$ inversion centers? First of all, if there is an odd number of inversion centers ($N_I=$odd), the inversion eigenvalue is not a gauge invariant quantity since the symmetry operations on a single spinon can always be followed by an arbitrary gauge transformation. In the case when $N_I=$even, if the inversion centers are not related by any other symmetry, again they can acquire extra gauge transformations independently under symmetry operations, and again the inversion eigenvalue is not a topological invariant.

If an even number of inversion centers are related by symmetry, on the other hand, one can compute their contribution to $I_s$ eigenvalue from parton PSGs in a universal manner. Without loss of generality, let's consider a pair of inversion centers related by certain crystal symmetry $\hat{P}$ (e.g. it could be a mirror reflection or a translation on a finite cylinder), this spinon pair operator transforms under inversion $\hat{I}_s$ as
\bea
&\notag\pso{I_s}\big[f_{{\bf r},\sigma}\cdot\pso{P}f_{{\bf r},\sigma}\psoinv{P}\big](\pso{I_s})^{-1}\\
&\notag=\pso{I_s}f_{{\bf r},\sigma}\psoinv{I_s}\cdot\pso{I_s}\pso{P}f_{{\bf r},\sigma}\psoinv{P}\psoinv{I_s}\\
&\notag=\eta^f_{I_s,P}\pso{I_s}f_{{\bf r},\sigma}\psoinv{I_s}\cdot\pso{P}\pso{I_s}f_{{\bf r},\sigma}\psoinv{I_s}\psoinv{P}\\
&=\eta^f_{I_s,P}\eta^f_{I_s}~\big[f_{{\bf r},\sigma}\cdot\pso{P}f_{{\bf r},\sigma}\psoinv{P}\big].
\eea
where we defined spinon PSGs
\bea
&\pso{I_s}\pso{P}f_{{\bf r},\sigma}\psoinv{P}\psoinv{I_s}=\\
&\notag\eta^f_{I_s,P}~\pso{P}\pso{I_s}f_{{\bf r},\sigma}\psoinv{I_s}\psoinv{P},\\
&(\pso{I_s})^2f_{{\bf r},\sigma}(\pso{I_s})^{-2}=\eta^f_{I_s}~f_{{\bf r},\sigma}.
\eea

Consequently, for a $N_s$-site lattice with $N_I$ inversion centers ($N_s,N_I=$even) which are pairwise related by crystal symmetry $\hat{P}$, the inversion eigenvalue of projected wavefunction (\ref{eq:Gproj}) is given in terms of spinon PSGs by
\bea
&\notag Q_{I_s}(N_s,N_I)=(-\eta^f_{I_s})^{(N_s-N_I)/2}(\eta^f_{I_s}\eta^f_{I_s,P})^{N_I/2}\\
&=(-\eta^f_{I_s})^{N_s/2}(-\eta^f_{I_s,P})^{N_I/2}.\label{eq:inversion:site:f}
\eea
for Abrikosov fermions and
\bea\label{eq:inversion:site:b}
& Q_{I_s}(N_s,N_I)=(\eta^b_{I_s})^{N_s/2}(\eta^b_{I_s,P})^{N_I/2}.
\eea
for Schwinger bosons.

A crucial point is that the PSG $\eta_{I_s,P}$ can depend on the boundary condition of the cylinder (for example, if $P = T_y^{L/2}$).

\subsubsection{Eigenvalue of mirror reflection operator $R$}

The eigenvalues of mirror reflection operator $R$ can be computed completely in parallel to the case of inversion symmetry as discussed previously. Again let's assume $N_R$ lattice sites lie on the mirror reflection axis on a $N_s$-site lattice. As argued earlier, only when $N_R$ is even and these $N_R$ sites are related to each other by other crystal symmetries, will the $R$ eigenvalue be a topological invariant that is fully determined by parton PSGs. Let's assume these $N_R$ sites are exchanged in pairs by crystal symmetry $\hat{P}$. Similar to the case of inversion symmetry $I_s$ we can compute the reflection eigenvalue of projected wavefunction (\ref{eq:Gproj}) as
\bea
Q_R(N_s,N_R)=(-\eta^f_{R})^{N_s/2}(-\eta^f_{R,P})^{N_R/2}.\label{eq:reflection:site:f}
\eea
for Abrikosov fermions and
\bea
Q_R(N_s,N_R)=(\eta^b_{R})^{N_s/2}(\eta^b_{R,P})^{N_R/2}.\label{eq:reflection:site:b}
\eea
for Schwinger bosons. The parton PSGs are defined as
\bea
&\pso{R}\pso{P}f_{{\bf r},\sigma}\psoinv{P}\psoinv{R}=\\
&\notag\eta^f_{R,P}\pso{P}\pso{R}f_{{\bf r},\sigma}\psoinv{R}\psoinv{P},\\
&(\pso{R})^2f_{{\bf r},\sigma}(\pso{R})^{-2}=\eta^f_{R}f_{{\bf r},\sigma}.
\eea
for Abrikosov fermions and similarly for Schwinger bosons.
	
\subsubsection{Unifying bosonic and fermionic PSGs}

If an Abrikosov-fermion state and a Schwinger-boson state describes the same $Z_2$ spin liquid state, their symmetry quantum numbers on any finite lattice must be the same for arbitrary crystal symmetries. Therefore from the eigenvalues of inversion and reflection symmetries summarized previously, we can achieve a unification of Abrikosov-fermion and Schwinger-boson representation: i.e. their PSGs must satisfy the following correspondence:
\bea
&-\eta_{I_p}^f=\eta_{I_p}^b;\\
&-\eta_{I_s}^f=\eta_{I_s}^b;\\
&-\eta_{I_s,P}^f=\eta_{I_s,P}^b,\\
&\notag\forall~\text{crystal symmetry}~P~\text{satisfying}~PI_s=I_sP;\\
&-\eta_{R}^f=\eta_{R}^b;\\
&-\eta_{R,P}^f=\eta_{R,P}^b,\\
&\notag\forall~\text{crystal symmetry}~P~\text{satisfying}~PR=RP.
\eea
These relations are in agreement with our conclusions based on the relative quantum numbers.
In the next section we'll establish the correspondence between Schwinger-boson and Abrikosov-fermion representations for those PSGs concerning time reversal symmetry. This is achieved by relating the 2D parton PSGs to 1D SPT invariants by considering projected wavefunctions on a thin but long cylinder.

\section{Dimensional reduction and entanglement signatures}
	A 2D model defined on a cylinder can be viewed as a 1D system by grouping  one ring of the cylinder into a single super-site.
This point of view is useful because the interplay of symmetry, topology and entanglement has been completely understood in 1D through the recent classification of 1D SPT phases. \cite{FidkowskiKitaev2011, PollmannTurnerBergOshikawa2010, ChenGuWen2011, Schuch}
In this section we explain how  2D SET order manifests itself as 1D SPT order under this dimensional reduction.
In particular, we find the Z$_2$ PSG relations have a one-to-one correspondence with the U(1) projective representations that classify 1D SPT phases.
While generally U(1) projective representations are a coarse-grained version of Z$_2$ projective representations, space-group symmetries actually have an \emph{anti}-unitary character under the dimensional reduction, and for this special case the correspondence becomes one-to-one.

\subsection{A review of 1D SPT phases.}
While the classification of 1D SPTs can be discussed in terms of Schmidt decomposition, the most compact treatment uses the formalism of matrix product states.
We refer to previous works for a more detailed review. \cite{PollmannTurnerBergOshikawa2010, ChenGuWen2011, Schuch,FidkowskiKitaev2011}

\subsubsection{Matrix product states}
Let $\ket{ j_n }$ span the local Hilbert spaces of a spin chain with sites at $n$.
A MPS $\ket{\Psi}$ is characterized by a sequence of rank-3 tensors $ \{ \Gamma^{j_n}_{\alpha_n \alpha_{n+1} } \}$ and rank-1 vectors $ \{ s_{\alpha_n} \} $ through the ansatz
\begin{align}
\braket{ \{j_n\} | \Psi} =  \sum^{\chi_n}_{ \alpha_n  = 1}   \prod_n  s_{\alpha_n} \Gamma^{j_n}_{ \alpha_n \alpha_{n+1} }.
\end{align}	
The indices $\alpha_n$ which are summed over are called the auxillary indices, with  dimension $\chi_n$.
We have assumed the MPS is in the `canonical form,' which means that each $s_{\alpha_n}$ is the set of Schmidt weights for a bipartiton of the system between sites $n-1, n$.

The MPS ansatz includes both finite and infinite spin chains.
In the finite case with $L$ sites, $\chi_1 = \chi_{L+1} = 1$.
In the infinite case with a unit cell of length $L$, we can always choose the tensors to share this unit cell: $\Gamma^{j_{n+L}}_{\alpha_{n+L} \alpha_{n+1 + L} } = \Gamma^{j_n}_{\alpha_n \alpha_{n+1} }$ and likewise for $s$ and $\chi$.

\subsubsection{Onsite symmetries}
If a spin chain is invariant under an onsite symmetry $g \in G$  (e.g. a spin rotation) , it is natural to ask how the symmetry is encoded in the tensors $\Gamma$.
The representation of the onsite symmetry decomposes into its action on each site, $\hat{g} = \otimes_n \hat{g}_n$.
For notational simplicity, we will drop the site index $n$.
An MPS is symmetric under $g$ if and only if the $\Gamma$ and $s$ transform as \cite{PerezGarcia2008}
\begin{align}
\label{eq:mps_onsite}
\sum_k g_{j k} \Gamma^{k}_{\alpha \beta} &=  r_g \,\sum_{\alpha', \beta'} U_{g; \alpha \alpha' } \Gamma^{j}_{\alpha' \beta'} U^\dagger_{g; \beta' \beta } \\
U_{g; \alpha \beta} s_\beta &= s_\alpha  U_{g; \alpha \beta}
\end{align}
where the $U_g$ are unitary matrices and $r_g \in$ U(1).

	The phases $r_g$ form a U(1) representation of the group, and encode the $g$-charge per unit length.
But for the unitaries $U_h$ there is another possibility: $U_g$ may be a \emph{projective} representation, meaning that the requirements of a group representation are satisfied only up to U(1) phases.
This subtlety arises because the phase of $U_g$  isn't fixed by the transformation law of Eq.~\ref{eq:mps_onsite}.
Arbitrarily fixing the phase of each $U_g$, the U(1) phases $\omega$ are encoded in the relations
\begin{align}
U_g U_h = \omega(g, h) U_{gh}, \quad \omega(g, h) \in \mbox{U(1)}.
\label{eq:projrep}
\end{align}
The phases $\omega$ are called cocycles, or the factor set.
If we consider the phase ambiguity $U_g \to \theta(g) U_g$, $\theta(g) \in $U(1) to be a `gauge transformation,' we see that $\omega$ is not gauge invariant.
The classification of gauge-inequivalent $\omega$ is given by the 2nd group cohomology  $[\omega] \in \mathcal{H}^2[G, U(1)]$, resulting in a classification of 1D  phases symmetric under $G$.
Note that for symmetry group $G = \mathbb{Z}_N$, there are no projective representations, and hence no 1D SPT phases.

An important physical signature of an SPT phase under onsite $G$ is degenerate edge states.
It can be shown there are edge states which transform under $G$ with the same projective representation $[\omega]$ as the $U_g$; if $[\omega]$ is non-trivial, the projective representation must be multi-dimensional, implying a degeneracy.

\subsubsection{Time-reversal symmetry}
The transformation laws are modified for time-reversal  $\mathcal{T} = \otimes_n \hat{u}_\mathcal{T} K$ because of the complex conjugation $K$.
Similar to before, the MPS can be taken to transform as
\begin{align}
\label{eq:mps_tr}
\sum_k u_{\mathcal{T}; j k} \bar{\Gamma}^{k}_{\alpha \beta} &=   \sum_{\alpha', \beta'} U_{\mathcal{T}; \alpha \alpha' } \Gamma^{j}_{\alpha' \beta'} U^\dagger_{\mathcal{T}; \beta' \beta } \\
U_{\mathcal{T}; \alpha \beta} s_\beta &= s_\alpha  U_{\mathcal{T}; \alpha \beta}
\end{align}
(note we can remove $r_{\mathcal{T}}$ by redefining the U(1) phase of the state).
But for an anti-unitary symmetry like time-reversal $\mathcal{T}$, the  projective relations are modified to
\begin{align}
U_{\mathcal{T}} U^\ast_h &= \omega(\mathcal{T}, h) U_{\mathcal{T}h}, \\
 U_h U_{\mathcal{T}} &= \omega(h, \mathcal{T} ) U_{h \mathcal{T}} \quad \omega \in \mbox{U(1)}
\label{eq:projrepT}
\end{align}
In contrast to an onsite $G = \mathbb{Z}_2$, for the anti-unitary time-reversal $(G = \mathbb{Z}_2^T)$ we can form the gauge-invariant relation
\begin{align}
\gamma_{\mathcal{T}} \equiv U_{\mathcal{T}} U^\ast_{\mathcal{T}}   = \pm 1.
\end{align}
This $\gamma_{\mathcal{T}}$ gives  a $\mathbb{Z}_2$ SPT classification.

There is also an interplay between the other symmetries and $\mathcal{T}$.
If $G$ contains $H \times \mathbb{Z}_2^T \subset G$ as a subgroup, where $H$ is onsite, for $h \in H$ there is a projective relation
\begin{align}
U_\mathcal{T} U^\ast_h U^{-1}_{\mathcal{T}} = \frac{\omega(\mathcal{T}, h) }{\omega(h, \mathcal{T}) }  U_h
\end{align}
For our purposes it will be sufficient to understand $A = \mathbb{Z}_2$; since $a \mathcal{T}$ is itself a unitary $\mathbb{Z}_2$ symmetry we can form the gauge-invariant relation
\begin{align}
\label{eq:hT_inv}
U_{h \mathcal{T}} U^\ast_{h \mathcal{T}}  = \gamma_{h \mathcal{T}} = \pm 1, \quad ( \mbox{when } \, h^2 = 1),
\end{align}
an additional $\mathbb{Z}_2$ invariant.
Alternatively, we have the relation
\begin{align}
(U_h^2) ( U_\mathcal{T} U^\ast_h U^{-1}_{\mathcal{T}} U_h^{-1}) = \gamma_{h \mathcal{T}} \gamma_{\mathcal{T}}
\end{align}

\subsubsection{Reflection symmetry}
Finally, consider a reflection $R$ which spatially inverts the 1D chain, $R \ket{j_n} = u_{R; j_n, k_{-n}} \ket{k_{-n}}$.
The unitary matrix $u_{R}$ encodes any internal rotation in the definition of the reflection.
The transformation law is \cite{PollmannTurnerBergOshikawa2010}
\begin{align}
\label{eq:mps_inv}
\sum_k u_{R; j k} {\left( \Gamma^T \right ) }^{k}_{\alpha \beta} &=  r_{R} \sum_{\alpha', \beta'} U_{R; \alpha \alpha' } \Gamma^{j}_{\alpha' \beta'} U^\dagger_{R; \beta' \beta } \\
U_{R; \alpha \beta} s_\beta &= s_\alpha  U_{R; \alpha \beta}
\end{align}
where $r_{R} = \pm 1$ provides the first $\mathbb{Z}_2$ invariant, the `parity per unit length.'
Somewhat surprisingly, combining this transformation law with those of an onsite $h$ we find the projective relations are the same as those of a \emph{anti}-unitary symmetry:
\begin{align}
U_{R} U^\ast_h &= \omega(R, h) U_{Rh}, \\
U_h U_{R} &= \omega(h, R ) U_{h R} \quad \omega \in \mbox{U(1)}.
\label{eq:projrepI}
\end{align}
The origin of this similarity is that transposition $T$  and complex conjugation behave analogously when acting on the unitary $U_h$.
This point is important, as  it implies inversion has the same $\mathbb{Z}_2$ invariants as time-reversal $\mathcal{T}$:
\begin{align}
U_{R} U^\ast_{R} &= \gamma_{R} = \pm 1 \\
U_{h R} U^\ast_{h R} &= \gamma_{h R} = \pm 1 \quad ( \mbox{when } \, h^2 = 1)  \\
(U_h^2) ( U_R U^\ast_h U^{-1}_{R} U_h^{-1}) &= \gamma_{h R} \gamma_R = \pm 1 
\end{align}

\subsection{Identification of 1D SPT order and the space-group PSGs}
	Earlier we argued that space-group quantum numbers are topological invariants in the presence of reflection symmetries, and calculated these quantum numbers using the PSG.
We now show that under the dimensional reduction these quantum numbers can be calculated from the 1D SPT invariants $r_{R}, \gamma_{R},  \gamma_{h R}$.
This clarifies the origin of their stability, since 1D SPT phases are robust  in the presence of symmetries, and provides a dictionary between the 2D SET and 1D SPT order.

To equate the 2D SET order with the 1D SPT invariants, we must compute the quantum numbers of a finite chain given the \emph{infinite} chain 1D SPT invariants.
For notational simplicity, we will assume the dimensional reduction results in a one-site 1D unit cell.
The notation is more complex for the Kagome model, since under the dimensional reduction it is more convenient to use a unit cell of two, so we delay the discussion of this case.
Since the system is a 2D cylinder, in addition to a reflection $R$ which exchanges the edges of the chain, there may be an orthogonal reflection $R'$ which does not exchange the edges, but instead behaves as an `onsite' symmetry under the dimensional reduction.
We combine these two reflections to form a 180 degree inversion $I = R R'$, which again exchanges the edges.
Consequently there are $\mathbb{Z}_2$ 1D SPT invariants $\gamma_R, \gamma_I = \gamma_{R R'}$.

For any $R / I $ symmetric state on a finite chain with $L$ sites ($L$-odd implies a site-centered inversion) it can be proven that, independent of any details of the edge,  the quantum numbers of the chain are
\begin{align}
Q_R(L) &= \gamma_R \left( r_R \right)^L \\
Q_I(L) &= \gamma_I \left( r_I \right)^L
\end{align}
where $\gamma, r$ are the 1D SPT invariants of the bulk phase.

To relate the 1D SPT invariants to the PSG, fix the transverse geometry $\Lambda_\perp$ of the cylinder (such as the circumference), and find the four ground states of the infinite cylinder. 
Each topological sector $a$ has 1D SPT invariants $\gamma_U(\Lambda_\perp, a), r_U(\Lambda_\perp, a)$.

Following our earlier discussion, we first compute the ratio of quantum numbers after threading a spinon $b/f$:
\begin{align}
Q_U^{(b/f)} = \frac{Q_U( \Lambda, b/f \cdot a)}{Q_U( \Lambda, a)} = \frac{\gamma_U(\Lambda_\perp, b/f\cdot a)}{\gamma_U(\Lambda_\perp, a)} (\frac{r_U(\Lambda_\perp, b/f\cdot a) }{r_U(\Lambda_\perp, a)})^L
\end{align}
Since the result is \emph{independent} of the geometry, we have the following:
\begin{align}
\label{eq:SPT_Qs}
Q_U^{(b/f)} &= \frac{\gamma_U(\Lambda_\perp, b/f\cdot a)}{\gamma_U(\Lambda_\perp, a)}  \\
r_U(\Lambda_\perp, b/f\cdot a)  &= r_U(\Lambda_\perp, a)
\end{align}
We find that \textbf{the ratio of 1D SPT invariants $\gamma_U$ between the topological sectors of an infinite cylinder reveals the PSG relation $Q_U^{(b/f)}$}.
Furthermore, the parity per unit length is unchanged by threading a spinon.

We then compute the ratio of quantum numbers after threading a vison $v$:
\begin{align}
Q_U^{(v)} = \frac{Q_U( \Lambda, v \cdot a)}{Q_U( \Lambda, a)} = \frac{\gamma_U(\Lambda_\perp, v \cdot a)}{\gamma_U(\Lambda_\perp, a)} (\frac{r_U(\Lambda_\perp, v \cdot a) }{r_U(\Lambda_\perp, a)})^L
\end{align}
When $U$ is not site-centered,  our earlier discussion found that $Q_U^{(v)} = 1$, so by comparison we expect
\begin{align}
\gamma_U(\Lambda_\perp, v \cdot a) &= \gamma_U(\Lambda_\perp, a) \\
r_U(\Lambda_\perp, v \cdot a) &= r_U(\Lambda_\perp, a), \quad U \in \{ R, I_{h/p} \}.
\end{align}
On the other hand, the site-centered inversion $I_s$ requires odd $L$, and from $Q^{(v)}_{I_s} = -1$ we find
\begin{align}
\frac{r_{I_b}(\Lambda_\perp, v \cdot a) }{r_{I_b}(\Lambda_\perp, a)} = -1
\end{align}

\subsubsection{Translating between the Z$_2$ PSG and  the 1D SPT U(1) PSG}
	The bosonic Z$_2$ PSG is encapsulated by the projective relations
\begin{align}
X \cdot Y = \omega^b(X, Y) (XY)  
\end{align}
where $\omega^b \in \mathbb{Z}_2$.
The  1D SPT relations are similar, but $\omega \in U(1)$, which is in general a much coarser classification as there are more phase ambiguities.
Yet we have shown that the 1D SPT relations recover the 2D PSG. 
So how do the 2D PSG relations `descend' to the 1D SPT relations?
The key point is that if the symmetry $R$ exchanges the edges of the cylinder, we can transcribe the 2D PSG relations into 1D SPT relations if we remember that reflection / inversion becomes anti-unitary in the 1D SPT realization:
\begin{align}
R^2 = \eta^b_R  \Rightarrow U_R U^\ast_R = \gamma_R
\end{align}
Due to the anti-unitary nature, the U(1) phase ambiguity does not affect the robustness of $\gamma_R$.
In general, \textbf{the 2D PSG relations descend to 1D SPT relations if we treat edge-exchanging symmetries as anti-unitary}.
This explains the equality of the quantum numbers found in Eq.~\eqref{eq:SPT_Qs}.

\subsection{Identification of 1D SPT order and the $R T R^{-1} T^{-1}$ PSGs}
\label{sec:1D_RT}
	In the fermionic parton construction there is an additional fermion PSG associated with the interplay of time-reversal $T$ and a reflection $R$:
\begin{align}
\eta^f_{R T} =  R^{-1} T^{-1} R T 
\end{align}
Viewing $R$ as on-site, under the dimensional reduction we have a similar 1D relation
\begin{align}
U_R U_T = \frac{\omega(R, T)}{\omega(T, R)} U_T U^\ast_R
\end{align}
But  $\omega(R, T)$ is \emph{not} U(1) gauge invariant.
Instead, we may consider the 1D SPT invariant defined by Eq.~\eqref{eq:hT_inv}.
Comparing  the 2D and 1D SPT PSG relations,
\begin{align}
( R^2) (R^{-1} T^{-1} R T ) &= \eta^f_R \eta^f_{RT} \quad & \mbox{2D PSG}\\
(U_R)^2 (U_R^{-1} U_T  U^\ast_R U_T^{-1})  &=  \gamma_{RT} \gamma_T \quad &\mbox{1D SPT} 
\end{align}
we obtain the following 2D PSG to 1D SPT reduction:
\begin{align}
\eta^f_R \eta^f_{RT} \rightarrow  \gamma_{RT} \gamma_T .
\end{align}
This identification can be verified by checking for the physical signature of $\eta^f_R \eta^f_{RT}  = -1$. 
The fermionic spinon always has a 2-fold degeneracy associated with it's Kramer's degeneracy $\eta_T^f = -1$, but when $\eta^f_R \eta^f_{RT}  = -1$  the spinon excitation is at least four-fold degenerate, as this is the minimal dimension of these projective relations.
The 1D SPT relation $ \gamma_T  = - \gamma_{RT}$ implies \emph{precisely} this additional two-fold degeneracy.

In summary, we can determine $\eta^f_R \eta^f_{RT}$ by computing the change in the 1D SPT invariant  $\gamma_T  \gamma_{RT}$ (with $R$ onsite) when the topological flux changes by $f$.
In Ref.~\onlinecite{LuAV2013} it argued that when $\eta^f_R = -1$,  $\eta^f_{RT}  = 1$ (where $R = \sigma$ in the notation of the Kagome model), there \emph{may} be gapless edge modes protected by a non-trivial \emph{vison} PSG $ R^{-1} T^{-1} R T  = -1$ (associated with topological superconductivity in the fermionic mean-field ansatz). 
Presumably the vison PSG can also be computed from the change in the 1D SPT invariant  $\gamma_{RT}$ ( $\gamma_T = 1$ for the vison) when threading a vison.

\section{Intrinsic topological order: detecting the topological flux}
It is important to have a method for measuring the topological flux $a$ of an MES  independent of the SET order;
in the finite cylinder, we must detect whether a vison $v$ threads the cylinder, and for a torus / infinite cylinder we must distinguish between all of $1, b, v, f$.
It has previously been shown that the topological $S$ and $T$ matrices can be calculated from the MES on both the torus and infinite cylinder, which can then be used to  label the MES.\cite{Zhang2012}
In practice this is not so simple as all four-sectors must be found on a finite circumference cylinder, which is frustrated by finite size effects which lead to a sizable splitting of the topological degeneracy.
However, for a $Z_2$ spin-liquid we find there is a simpler procedure to uniquely label the MES individually.

Note that an $S=1/2$ model must have finite topological-flux per unit cell \cite{MikeAV2014}, so the topological flux through an entanglement cut depends on the location of the cut.
So to simplify the discussion, we restrict  to even circumferences $L_y \in 2 \mathbb{Z}$, which contain an even number of $S = 1/2$ within each ring of the cylinder, and always consider 'vertical' entanglement cuts which  lie between 1D  super-sites under the 2D to 1D dimensional reduction.

\subsection{ $ \{ \mathds{1}, v \}$ vs $ \{ b, f \}$}
The first distinction detects the fractional $S = 1/2$ spin carried by the spinons  $b$ and $f$.
Under the dimensional reduction, an even circumference cylinder is an SO(3) invariant integer-spin chain, which has a $\mathbb{Z}_2$ 1D SPT classification associated with the emergence of two-fold degenerate $S = 1/2$ edge states (protected either by time-reversal or SO(3) ). 
The simplest non-trivial example is the $S = 1$ AKLT state.
In 2D, when topological flux $b$ or $f$ terminates at the edge of the cylinder it also produces a spinon excitation near the edge carrying $S = 1/2$.
If the system is SO(3) symmetric, this emergent edge spin carries a two-fold degeneracy.
Hence under the dimension reduction, the $b, f$ sectors are non-trivial 1D SPT states under SO(3), while $\mathds{1}, v$ are trivial.

We conclude that the $\mathds{1}, v$ sectors will have an entanglement spectrum that transforms under integer representations of SO(3), and hence will have singlets in the entanglement spectrum, while the entanglement spectrum of the $b, f$ sectors will transform under half-integral representations of SO(3), leading to a minimum two-fold  entanglement degeneracy.

\subsection{ $ \mathds{1}$ vs $v$}
	In an  SO(3) symmetric spin liquid, threading a vison $v$ through the system is topologically equivalent to  threading $2 \pi$ - flux with respect to $S^z$ spin-rotations.
To distinguish between the $\mathds{1}$ and $v$ sectors we can detect the change in momentum induced by the flux threading.
Viewing a cylinder of length $L_x$ and circumference $L_y$ as a spin chain on a periodic ring of length $L_y$, the unit cell of the chain contains integral / half-integral spin when $L_x$ is even / odd.
In the half-integral case, threading flux with respect to $S^z$ spin rotations is known to increase the $y$-momentum by $e^{i \pi} = -1$.\cite{LSM1961}
In the integral case, the  threading flux will not change the momentum.

We  conclude that threading a vison through the system will increment the $y$-momentum  by $\pi$ when $L_x$ is odd, and by $0$ if $L_x$ is even.
This increment is simply the $y$-momentum per unit length $\eta^{\mathds{1} / v}_{xy}$ introduced in Sec. \ref{sec:translationQN}.
Hence for the vacuum, $\eta^{\mathds{1}}_{xy} = 1$, while for the vison, $\eta^{v}_{xy} = -1$.
This result is straightforward to check in any parton construction.

We know $\eta^b_{xy} = - \eta^f_{xy}$, since they differ by vison insertion,  but \emph{which} of the two carries $\eta^a_{xy} = -1$ will in fact depend on the PSG relation $(T_x T_y)^a = \eta^{a}_{xy} (T_y T_x)^a$.

\subsection{ $b$ vs $f$}
	A fermionic anyon has topological spin $\theta_f = -1$, which we can use to distinguish between the $b$ and $f$ spinons.
We will show that the fermion's topological spin is encoded in an additional 2x degeneracy in the entanglement spectrum (ES); combined with the 2x spin degeneracy, the fermionic spinon ES if 4x degenerate.
Intuitively, the ES of the fermionic sector should have anti-periodic boundary conditions, meaning that the momenta $k$ are quantized as $k \in \frac{2 \pi}{L_y} (\mathbb{Z} + \tfrac{1}{2}) $.
With either time-reversal or reflection symmetry, the momenta $k$ and $-k$ will be degenerate, so there is a 2x degeneracy.
We will show this 2x degeneracy arises from non-trivial 1D SPT order under a combination of $\mathbb{Z}_{L_y}$ rotational symmetry and either reflection or time-reversal.

\subsubsection{Review of momentum polarization}
Momentum polarization is a procedure to detect topological spin using a translation $T_y$ that rotates a cylinder of circumference $L_y$. \cite{ZaletelMongPollmann2013, TuZhangQi}
To review, each left Schmidt state $\ket{\alpha}$ of the Schmidt decomposition $\{ e^{-E_\alpha}, \ket{\alpha} \}$ can be assigned definite momentum $e^{i k_\alpha}$, meaning that $T_y \ket{\alpha} = e^{i k_\alpha} \ket{\alpha}$.
In a convention in which $k_\alpha \in \frac{2 \pi}{L_y} \mathbb{Z}$, the momentum polarization is
\begin{align}
\left( \sum_\alpha e^{-E_\alpha + i k_\alpha } \right)^{L_y} &= \mathcal{T}_a \, e^{- (\alpha - i \eta_H) L_y^2}  + \mathcal{O}(e^{-L_y/\xi}), \\
\mathcal{T}_a &= e^{ 2 \pi i (h_a - c / 24) }.
\end{align}
Here $\alpha$ is non-universal real constant related to the area-law fluctuation of momentum of the cut; $\eta_H$ is the `Hall-viscosity' (which isn't quantized on a lattice), and $\mathcal{T}_a = e^{ 2 \pi i (h_a - c / 24) }$ is the desired entry of the modular $\mathcal{T}$-matrix.

The momentum polarization is \emph{not} necessarily a 1D SPT invariant, since there are no 1D SPTs associated with an onsite $\mathbb{Z}_{L_y}$ symmetry.
As a consequence, generally the momentum polarization only becomes quantized in the $L_y \to \infty$ limit, and a scaling analysis is required.

\subsubsection{A 1D SPT invariant for detecting fermionic topological flux }
However, in the presence of a mirror reflection $y \leftrightarrow -y$ or time reversal, we can prove that the momentum polarization is a $\mathbb{Z}_2$ 1D SPT invariant that detects whether the topological flux is bosonic or fermionic.
Consider a cylinder of even circumference $L_y$ with a mirror reflection $R_y: (x, y) \to (x, -y)$ that acts as an onsite symmetry in the 1D picture (the result for time reversal is analogous).
When acting on the entanglement spectrum, a $\pi$-rotation (translation by $L_y/2$) may anti-commute with the inversion $R_y$:
\begin{align}
U_{R_y} \, (U_{T_y})^{{L_y}/2} \,  U^{-1}_{R_y} = (-1)^F (U_{T_y})^{{L_y}/2}, \quad F = 0, 1
\label{eq:Fdef}
\end{align}
giving a 1D $\mathbb{Z}_2$ invariant $F = 0, 1$. To show $F$ is a 1D SPT invariant, note the symmetry group generated by $T_y, R_y$ is $G = \mathbb{Z}_{L_y} \rtimes \mathbb{Z}_2$ (for even $L_y$).
The cohomology classification is
\begin{align}
\mathcal{H}^2[ \mathbb{Z}_{L_y} \rtimes \mathbb{Z}_2 , U(1)] = \mathbb{Z}_2.
\end{align}
The relation $F$ of Eq.~\eqref{eq:Fdef} is a gauge invariant, so must label these two possibilities.

To relate the SPT invariant $F$ to the anti-periodic boundary conditions of the entanglement spectrum, suppose we redefine the phase of $U_{T_y}$ to ensure the expected relation $U^{-1}_{R_y} U_{T_y} U_{R_y} =  (U_{T_y})^{-1}$.
With this gauge choice Eq. \eqref{eq:Fdef} requires $\left( U_{T_y} \right)^{L_y} = (-1)^F$.
For $F = 1$, the diagonal basis $U_{T_y} = e^{i k_\alpha}$ requires $k_\alpha \in \frac{2 \pi}{L_y} ( \mathbb{Z} + \tfrac{1}{2})$.

We note that the most general commutation relation is
\begin{align}
U_{R_y} U_{T_y} U^{-1}_{R_y} = \gamma_F \, U^{-1}_{T_y}
\end{align}
with $\gamma_F^{{L_y}/2} = (-1)^F$.

\subsubsection{Quantization of momentum polarization by the 1D SPT invariant}
We now prove that the momentum polarization is quantized to be $(-1)^F$; this confirms the interpretation that $F$ detects the topological spin $\theta_f = -1$ of the fermionic spinon.

Since $U_{R_y}$ commutes with the entanglement spectrum $e^{-E_\alpha}$, we have
\begin{align}
\lambda &= \mbox{Tr}( e^{-E_\alpha} U_{T_y}) = \mbox{Tr}( e^{-E_\alpha} U^{-1}_{R_y} U_{T_y} U_{R_y})  \\
&= \mbox{Tr}( e^{-E_\alpha} \gamma_F U^{-1}_{T_y})= \gamma_F \mbox{Tr}( e^{-E_\alpha}  U^\ast_{T_y}) = \gamma_F \lambda^\ast
\end{align}
Using this relation, for even ${L_y}$ the momentum polarization is
\begin{align}
\lambda^{L_y} = \lambda^{{L_y}/2} (\gamma_F \lambda^\ast)^{{L_y}/2} = \gamma_F^{{L_y}/2} |\lambda|^{{L_y}} = (-1)^F |\lambda|^{{L_y}}.
\end{align}
So long at $|\lambda| \neq 0$, the momentum polarization is equivalent to the SPT invariant.
The SPT invariant is even more robust since it is well defined even when $|\lambda| = 0$.

\section{Detecting  SET order on the Kagome lattice using cylinder-DMRG}
	We now propose a procedure to determine the Kagome PSGs which is practically adapted to the constraints of cylinder DMRG.
The results of this analysis will be reported in a subsequent work.\cite{Zhu2015}

Following the notation of the earlier analysis, \cite{LuRanLee} within the fermionic parton construction we must determine the five invariants $\{ \eta_\sigma, \eta_{C_6}, \eta_{\sigma C_6}, \eta_{\sigma T}, \eta_{C_6 T}, \eta_{12} = \eta_{C_6} \eta_{\sigma C_6} \} $, combined with a possible $\nu_{x/y} \in \mathbb{Z}$ classification of any symmetry-protected  gapless edge states. \cite{LuAV2013}

\subsection{Finite DMRG}
	We first consider a technique  for finite length cylinders.
By finding the SO(3) invariant ground state after adding or removing an extra spin at each edge of the geometry, DMRG studies can reliably obtain two topological sectors that differ by threading a spinon through the bulk.
At the circumferences that can currently be well converged (such as YC8 and XC8), the other two topological sectors are not generally observed. 
Even without determining whether these two sectors are $1/v$ , $b/f$, the ratio of $R_x, R_y, I_h$ quantum numbers before and after adding the extra sites (i.e. a spinon) will determine $Q^{(b/f)}_{R_x, R_y, I_h}$, and referring to Table \ref{fig:KSL_Qa}, three of the PSG invariants.
Since the symmetries must be edge-exchanging, on the YC type cylinders we obtain $Q^{(b/f)}_{R_x}$ and  hexagon centered $Q^{(b/f)}_{I_h}$. 
On the XC type cylinders, we obtain $Q^{(b/f)}_{R_y}$ and  hexagon centered $Q^{(b/f)}_{I_h}$, the latter serving as a double check on the YC data.
There is a simple algorithm for measuring space-group quantum numbers in finite DMRG.\cite{Zhu2015}

To determine $\eta_{12}$ one must first determined whether the topological sector of the spinon is $b / f$. 
To distinguish $b / f$, we check if there is a 4-fold degeneracy in the entanglement spectrum as predicted by Eq.~\eqref{eq:Fdef}, which would imply the sector is $f$.
Knowing the sector $b / f$, we know the correct boundary condition for the parton ansatz, and following the techniques of Sec.~\ref{sec:GS_QN} we can predict the site-centered inversion $Q_{I_s}(\Lambda, b/f)$.
The details depend on the cylinder used, but the result always reveals $\eta_{12}$.

Once we know the topological sector of the sample, we can further check these results by comparing the \emph{absolute} quantum numbers under $R_x, R_y, I_{h/s}$ with the computation of Sec.~\ref{sec:GS_QN}; 
there are many different cases depending on the sample.

\subsection{Infinite DMRG}
	An infinitely long cylinder can be studied using iDMRG, which has certain numerical advantages due to the absence of edge effects and the reduced computational costs.
iDMRG also reliably finds two topological sectors which differ by a spinon.
Here we discuss only even-circumference cylinders.

As discussed, the $\mathbb{Z}_2$ 1D SPT invariants for SO(3) / $T$ determine which state has the spinon.
The fermionic 1D SPT invariant distinguishes between $b / f$.
The momentum per unit length $\eta_{xy}^{a}$ is trivial to compute in iDMRG, so we distinguish between sectors $1/v$ using $1 = \eta_{xy}^{1} = - \eta_{xy}^{v}$. 
The momentum per unit length of the spinon sector determines either $\eta_{12} = \eta_{xy}^f$ (if the sector is $f$) or  $-\eta_{12} = \eta_{xy}^b$ (if the sector is $b$). 

To measure the reflection PSGs, one can in principle detect the 1D SPT reflection invariants using established methods.\cite{PollmannTurner2012}
This can be a bit unwieldy in 2D DMRG, as the ordering of the DMRG `snake' breaks the reflection symmetries.
A simpler procedure is to generate a \emph{finite} cylinder wavefunction by projecting the left / right regions of the infinite cylinder  onto reflection related classical product states, leaving behind a finite segment of spins.\cite{Zhu2015} 
Regardless of projection used, the resulting state is a reflection symmetric finite cylinder wavefunction.
One can then  measure the space-group quantum numbers of the resulting finite cylinder wavefunction in order to determine $Q^{(b/f)}_{R_x, R_y, I_h}$

\subsection{Determining $R T R^{-1} T^{-1}$}
	As discussed in Sec.~\ref{sec:1D_RT},  the remaining invariants are related to the onsite 1D SPT invariants $\gamma_{R T}$ for $R = R_x, R_y$, which can be measured on YC and XC type cylinders respectively using known methods for detecting 1D SPTs. \cite{PollmannTurner2012}
The most obvious signature is the 4-fold degeneracy in the ES required to realize the projective relations $U_T U_T^\ast= \gamma_T = -1$ and $(U_R)^2 (U_R^{-1} U_T  U^\ast_R U_T^{-1})  = \gamma_T \gamma_{R T} = -1$.
The fermionic PSGs are related to the relative 1D SPT order $\gamma^{(f)}$ between the $1$ and $f$ sectors via
\begin{align}
\gamma^{(f)}_T \gamma^{(f)}_{R_x T} & = \eta_{\sigma T} \eta_{\sigma} \\
 \gamma^{(f)}_T \gamma^{(f)}_{R_y T} & = \eta_{\sigma T} \eta_{C_6 T} \eta_{\sigma} \eta_{\sigma C_6}
\end{align}
Of course $ \gamma^{(f)}_T = -1$, as it is a spinon.

If the DMRG obtains sectors which differ by $f$, we are done.
If DMRG obtains sectors which differ by $b$, the analysis depends on whether the vison has a non-trivial  $R T R^{-1} T^{-1}$ PSG.
If the vison PSG is trivial, there are no gapless edge states and the boson will have the same $R T R^{-1} T^{-1}$ PSG as the fermions, so the bosonic relative SPT order $\gamma^{(b)}$ again recovers the fermionic PSG invariants.
If the vison PSG is non-trivial, there are gapless edge states, and  $\eta_{\sigma T} \eta_{\sigma} = -1$ for the $R_x$ edge and $\eta_{\sigma T} \eta_{C_6 T} \eta_{\sigma} \eta_{\sigma C_6} = -1$ for the $R_y$ edge regardless.
	
\section{Conclusions}
We argued that the many body symmetry quantum numbers are a robust global property ideally suited to detecting  distinctions between SETs, and calculated these for several SETs described by slave particle mean field theories with different projective symmetry groups.  

More generally, the SETs may be diagnosed from the 1D SPT invariants in the cylinder geometry. These invariants for different topological sectors (which are labeled by the  quasiparticles) combine together in a way that reflects the fusion rules. In contrast, combining PSGs for a pair of quasiparticles to predict the PSG for the fusion product needs to be carefully considered in the case of internal symmetries.  

The knowledgeable reader may be puzzled by this correspondence between PSGs and 1D SPTs. The latter is determined by projective representations modulo a phase, or technically ${\mathcal H}^2(G,\,U(1))$, where $G$ is the symmetry group, while the former is a projective representation modulo Z$_2$, ${\mathcal H}^2(G,\,Z_2)$, for Z$_2$ topological order. That is, we represent group elements by matrices, whose product satisfies the group multiplication, up to either a U(1) phase or just an overall sign (Z$_2$) of the matrices . This is because the physically observable quantities are made by combining two identical quasiparticles and so we can only change the overall sign of the matrices. For example, if X is a Z$_2$ symmetry, then `half charge' of a quasiparticle corresponds to the PSG $X^2=-1$. However, this is not a 1D SPT invariant. How is this discrepancy reconciled?

The key observation is if inversion, $I$, or a equally a reflection is present, these act like anti-unitary symmetries  when restricted to the Schmidt states on one side of a bipartition. Thus, while $I^2=-1$ may again seem to be a PSG relation, regarding it  as an antiunitary symmetry turns it into a projective representation even with $U(1)$ phase factors and hence a 1D SPT invariant \cite{PollmannTurnerBergOshikawa2010, TurnerZhangAV}.  Similarly, for a global Z$_2$ symmetry, $X$, while by itself $X^2=-1$ does not produce a 1D SPT invariant, when combined with the effective antiunitary inversion symmetry  $\frac{XIX^{-1}I^{-1}}{X^2}$ is a 1D SPT invariant and the denominator is the fractional charge we are interested in. This type of reasoning has been repeatedly used in this work.

Our procedure is expected to be complete for Z$_2$ liquids, but for more complicated topological orders, such as say Z$_3$ topological order, there is a Z$_3$ invariant associated with a C$_3$ rotation symmetry. Taking into account certain subtleties, global C$_3$ quantum numbers detect this topological invariant, but the simplest cylinder dimensional reduction will not work. Extensions are left to future work. 

\section{Acknowledgements}
We thank Steve White, David Huse and  Zhenyue Zhu for a stimulating collaboration on related topics. We also thank Yang Qi and Liang Fu for alerting us to their recent preprint before submission, which overlaps with the present work\cite{QiFu}. AV was supported by NSF-DMR 1206728 and the Templeton Foundation. MZ was supported by NSF DMR-1206515 and the David \& Lucile Packard Foundation.
\bibliography{Refs}
\end{document}